\documentclass[pre,aps,twocolumn,showpacs,superscriptaddress]{revtex4-1}
\usepackage{graphicx}
\usepackage{amsmath,amssymb}
\usepackage{comment}
\usepackage{color}
\def\be{\begin{equation}}
\def\ee{\end{equation}}
\def\bea{\begin{eqnarray}}
\def\eea{\end{eqnarray}}

\def\nn{\nonumber}

\def\be{{\bm{\hat{e}}}}

\newcommand{\eref}[1]{Eq.~(\ref{#1})}%
\newcommand{\Eref}[1]{Equation~(\ref{#1})}%
\newcommand{\fref}[1]{Fig.~\ref{#1}} %
\newcommand{\Fref}[1]{Figure~\ref{#1}}%
\newcommand{\sref}[1]{Sec.~\ref{#1}}%
\begin{document}

\title{ Velocity Distribution of Driven Inelastic One-component Maxwell gas}

\author{V. V. Prasad} 
\affiliation{The Institute of Mathematical Sciences,
C.I.T. Campus, Taramani, Chennai-600113, India} 
\affiliation{Homi Bhabha National Institute, Training School Complex, Anushakti Nagar, Mumbai-400094, India}
\author{Dibyendu Das} 
\affiliation{Department of Physics, Indian Institute of Technology, Bombay, Powai, Mumbai-400076,India} 
\author{Sanjib Sabhapandit} 
\affiliation{Raman Research Institute, Bangalore - 560080, India} 
\author{R.Rajesh} 
\affiliation{The Institute of Mathematical Sciences,
C.I.T. Campus, Taramani, Chennai-600113, India} 
\affiliation{Homi Bhabha National Institute, Training School Complex, Anushakti Nagar, Mumbai-400094, India}
\date{\today}

\begin{abstract}
The nature of the velocity distribution of a driven granular gas, though well studied, is 
unknown as to whether it is universal or not, and if universal what it is.
We determine the tails of the steady state velocity distribution of a driven 
inelastic Maxwell gas, which is a simple model of a granular gas where the rate of collision 
between particles is independent of the separation as well as the relative velocity.  
We show that the steady state velocity distribution is 
non-universal and depends strongly on the nature of driving.
The asymptotic behavior of the velocity distribution are shown to be identical to that of a non-interacting model
where the collisions between particles are ignored. For diffusive driving, where collisions with the wall
are modelled by an additive noise,  the tails of the velocity distribution
is universal only if the noise distribution decays faster than exponential.
\end{abstract}
\maketitle

\section{Introduction}
\label{introduction}

Granular matter, constituted of particles that interact through inelastic collisions, exhibit diverse 
phenomena  such as cluster formation, jamming, phase separation, pattern formation,  static 
piles with intricate stress networks, etc.~\cite{Jaeger:96,Aranson:06, Goldhirsch:93, Li:03, Corwin:05}. 
Its ubiquity in nature and in industrial applications makes it important to understand how the 
macroscopically observed behavior of granular systems arises from the microscopic dynamics. A 
well studied macroscopic property is the velocity distribution of a dilute granular gas.  While several 
studies (see below) have shown that the inherent non-equilibrium nature  of the system, induced by inelasticity,
could result in a   non-Maxwellian velocity distribution, they fail to pinpoint whether the  
velocity distribution is universal, and if yes, what its form is. In this paper, we focus on 
the role of driving in determining the velocity distribution within a simplified model for
a granular gas, namely  the inelastic Maxwell model.

Dilute granular gases are of two kinds: freely cooling in which there is no input of energy~\cite{Haff:83,Brey:96,Esipov:97,Ben-naim:99,Ben-naim:00,Noije:98,Nie:02,Supravat:12, Pathak:14a,Pathak:14,shinde2007violation}, or driven, in which energy is injected at
a constant rate.
In the freely cooling granular gas, the velocity distribution
at different times $t$ has the form $P(v,t) \simeq v^{-1}_{rms} f(v/v_{rms})$, where $v$ is any of
the velocity components, $v_{rms}(t)$ is the time dependent root mean square velocity and $f$ is a scaling function.  
$v_{rms}(t)$ decreases in time as a power law $v_{rms}(t)\sim t^{-\theta}$.
To determine the behavior of $f$ for large argument, it was argued that the contributions to the tails of the velocity
distributions are from particles that do not undergo any collisions, implying an exponential 
decay  of $P(v,t)$ with  time $t$~\cite{Nie:02}. Thus, $f(x)\sim \exp(-a x^{1/\theta})$, or $P(v,t) \sim e^{-a v^{1/\theta} t}$ for 
large $v$. It is known that at initial times, the granular particles remain homogeneously distributed with $\theta=1$~\cite{Haff:83}, 
leading to $P(v,t)$ having an exponential decay in all dimensions. At late times they tend to cluster resulting in density inhomogeneities with current evidence suggesting $\theta=d/(d+2)$~ \cite{Ben-naim:99,Nie:02,Pathak:14}.

In dilute driven granular gases,  the focus of this paper, the system reaches a steady state where the energy lost 
in collisions is balanced by external driving. Several experiments, simulations and theoretical 
studies have focused on determining the steady state velocity distribution $P(v)$.
In experiments, driving is done either mechanically~\cite{Clement:91,Warr:95,Kudrolli:97,Olafsen:99,Losert:99,Kudrolli:00,Blair:01,Rouyer:00} through collision of the particles 
with vibrating wall of the container or by applying electric~\cite{Aranson:02} or magnetic fields~\cite{kohlstedt:05} 
on the  granular beads.
Almost all the experiments find the tails of $P(v)$ to be non-Maxwellian, and described by a  stretched exponential form $P(v) \sim \exp(-a v^\beta)$ for large $v$. Some of these experiments find $P(v)$ to be universal with $\beta=3/2$ for a wide range of parameters~\cite{Losert:99,Rouyer:00}.
In contrast, other experiments~\cite{Olafsen:99, Blair:01} find $P(v)$ to be non-universal with the exponent $\beta$ varying with the system parameters, sometimes even approaching a Gaussian distribution ($\beta=2$)~\cite{Olafsen:99}.  

In numerical simulations, driving is done either from the boundaries~\cite{Du:95,Esipov:97} which leads to clustering, or  homogeneously~\cite{Williams:96,Moon:01,Vanzon:04,Vanzon:05} within the bulk.
In simulations of a  granular gas in three dimensions, driven homogeneously by addition of white noise to the velocity ({\emph{diffusive driving}}),
it was observed that $\beta=3/2$ for large enough inelasticity~\cite{Moon:01}.
However, similar simulations of   a
bounded two dimensional  granular gases with diffusive driving
found a range of distributions in the steady state, with $\beta$ ranging from $0.7$ to $2$ as the parameters in the system
are varied~\cite{Vanzon:04,Vanzon:05}.
 
Theoretical approaches have been of two kinds:  kinetic theory, or by studying simple models which capture essential
physics but are analytically tractable. In kinetic theory~\cite{Brilliantov:04}, the Boltzmann equation describing the   
evolution of the distribution function is obtained by truncating the  BBGKY hierarchy by assuming product measure for
joint distribution functions. While it is difficult to solve this non-linear equation exactly, the deviation of the velocity
distribution from Gaussian can be expressed as a  perturbation expansion using Sonine polynomials~\cite{Goldshtein:95,Noije:98,Brilliantov:04,Dubey:13}.
This approach describes the velocity distribution near the typical velocities.
The tails of the distribution can be obtained by linearizing the Boltzmann equation~\cite{Noije:98,Montanero:00,Barrat:02}.
Notably, for granular gases with diffusive driving,  this leads to the prediction  $P(v)\simeq{\mathcal{C}}\exp(-b |v|^\beta)$ with
$\beta=3/2$  for large velocities, independent of the coefficient of restitution,
strongly suggesting  that the velocity distribution is
universal~\cite{Noije:98}.

The alternate theoretical  approach is to study simpler model like the inelastic Maxwell gas, in which spatial coordinates of the
particles are ignored and each pair of particles collide at constant rate~\cite{Ben-naim:00}.
In the freely cooling Maxwell gas, the velocity distribution decays as a power law with an exponent that depends on
dimension and coefficient of restitution~\cite{Baldassarri:02,Ernst:02_a, Krapivsky:02,Ben-naim:02}. 
In contrast, for a diffusively driven Maxwell gas,  in which collisions with the wall and modelled
by velocities being modified by an additive noise, it was shown that $P(v)$
has  a universal exponential tail ($\beta=1$) for all coefficients of restitution~\cite{Ernst:02,Antal:02}. 
However, it has been recently shown~\cite{Prasad:13,Prasad:14} that when the driving is diffusive, 
the velocity of the center of mass does a Brownian motion, and the total energy increases linearly with time at large times. Thus,
the system fails to reach a time-independent steady state, making the  results for diffusive driving valid only for intermediate times
when a pesudo-steady state might be assumed. This drawback may be overcome by modeling driving through collisions
with a wall, where the new velocity $v'$ of a particle colliding with a wall is given by $v'=-r_w v+\eta$, where $r_w$
is the coefficient of restitution for particle-wall collisions, and $\eta$ is uncorrelated noise representing the momentum transfer due to the wall~\cite{Prasad:13} (diffusive driving corresponds to $r_w=-1$). 
For this {\it dissipative} driving ($|r_w|<1$), the system reaches a steady state, and the velocity distribution was
shown to be Gaussian when $\eta$ is taken from a normalized Gaussian distribution~\cite{Prasad:13}. 
If $\eta$ is described by a Cauchy distribution, the steady state $P(v)$ is also 
a Cauchy distribution, but with a different parameter~\cite{Prasad:13}.

Thus,  while the velocity distribution for the freely cooling granular gas is universal and reasonably well understood, 
it has remained unclear whether the velocity distribution of a driven granular gas is universal. Also, if the velocity distribution
is non-Maxwellian, a clear physical picture for its origin is missing.
Intuitively, it would appear that the tails of the velocity distribution would be dominated by particles that have 
been recently driven and not undergone any collision henceforth. This would mean the $P(v)$ cannot decay faster 
than the distribution of the noise associated with the driving. 
If this reasoning is right, the noise statistics should play a crucial role
in determining the velocity distribution, making it non-universal. How sensitive is  $P(v)$ to the details of the driving?  
In particular, how does $P(v)$ behave for large $v$ for different 
noise distributions $\Phi(\eta)$?
We answer this question within the Maxwell model, both for dissipative driving ($0\leq r_w<1$) as well as
the pseudo steady state  for diffusive driving ($r_w=-1$). In particular, we show that the tail statistics are
determined by the noise distribution for dissipative driving. For the pseudo steady state in diffusive driving,
we find that the velocity distribution is universal if the noise distribution decays faster than exponential 
and determined by noise statistics if the noise distribution decays slower than exponential.

The rest of the paper is organized as follows. In \sref{model}
we define the Maxwell model and its dynamics more precisely. In \sref{steadystate general gaussian}
the steady state velocity distribution of the system are determined by studying its characteristic function as well as
the asymptotic behavior of ratios of successive moments.  In particular, we obtain the velocity distribution 
for a family of stretched exponential distributions for the noise. The results for 
dissipative driving may be found  in \sref{dissipative driving} and those for
diffusive driving in \sref{diffusive driving}. In \sref{section non-interacting}, 
the exact solution of the non-interacting problem is presented.  Section~\ref{conclusion} contains a summary 
and discussion of results.

\section{Driven Maxwell gas \label{model}}

Consider $N$ particles of unit mass. Each particle $i$ has a one-component
velocity $v_i$, $i=1,2, \ldots, N$. The particles undergo two-body collisions that 
conserve momentum but dissipate energy, such that when particles $i$ and $j$ collide, the 
post-collision velocities $v_i'$ and $v_j'$ are given in terms of the pre-collision velocities
$v_i$ and $v_j$ as: 
\begin{equation}
\begin{split}
\label{interparticle collision}
v_{i}' &=  \frac{(1-r)}{2} v_i + \frac{(1+r)}{2} v_{j},\\ 
v_{j}'&= \frac{(1+r)}{2} v_i + \frac{(1-r)}{2} v_{j},
\end{split}
\end{equation} 
where $r\in[0,1]$ is the coefficient of restitution. 
For energy-conserving elastic collisions, $r=1$.
In the Maxwell gas,  the rate of collision of a pair of particles is assumed to
be {\it independent of their spatial separation as well as their relative velocity.} 
These simplifying assumptions make the model more tractable as the spatial
coordinates of the particles may now be ignored.

The system is driven by input of energy, modeled by particles colliding with a  vibrating wall~\cite{Prasad:13}. 
If particle $i$ with velocity $v_i$ collides with the wall having velocity $V_w$, the new velocities
$v'_i$, $V'_w$ respectively, satisfy the relation $v_i'-V_w'=-r_w(v_i-V_w)$, where the parameter 
$r_w$ is the coefficient of restitution  for particle-wall collisions. 
 Since the wall  is much heavier than the particles, $ V_w'\approx V_w$, and hence
$v_i'=-r_w v_i+(1+r_w)V_w$. Since the motion of the wall is independent 
of the particles and the particle-wall collision times are random, it is reasonable to replace $(1+r_w)V_w$ 
by a random noise $\eta$ and the new velocity $v_i'$ is now given by~\cite{Prasad:13}, 
\begin{equation}
\label{driving sec2}
v_{i}' =-r_w v_i + \eta_i.
\end{equation}  
In this paper, we consider  a class of normalized stretched exponential
distributions for the noise $\eta$, 
\begin{align}
\Phi(\eta)=\frac{a^{\frac{1}{\gamma}}}{2\Gamma\left(1+\frac{1}{\gamma}\right)}\exp(-a|\eta|^\gamma)~
a,\gamma>0,
\label{ansatz noise pdf}
\end{align} 
characterized by the exponent $\gamma$.
Note that there is no apriori reason to assume that the noise is Gaussian as the noise is not averaged  over
many random kicks. 

The system is evolved in discrete time steps. At each step, a pair of particles are chosen at random and
with probability $p$ they collide according to Eq.~\eqref{interparticle collision}, and with probability $(1-p)$, 
they collide with the wall according to Eq.~\eqref{driving sec2}. We note that evolving the system 
in continuous time does not change the results obtained for the steady state.

We also note that  though the physical range of  $r_w$ is $[0,1]$, 
it is useful to mathematically extend its range to  $[-1,1]$. This makes it convenient to treat special  limiting cases in one
general framework.  For instance, when $r_w=-1$, the driving reduces to a random noise being added to the velocities, corresponding to diffusive driving. In this case, the system reaches a 
pseudo-steady state before energy starts increasing linearly with time for large times~\cite{Prasad:13,Prasad:14}. 
When $r_w \neq -1$, the system reaches a steady state that is independent of the initial conditions.
In the limit $r_w\to -1$, and rate of collisions with the wall going to infinity, the problem reduces to an Ornstein-Uhlenbeck
process~\cite{Prasad:14}. The case $r_w=1$ is also interesting.   When $r_w=1$, the structure 
of the equations obeyed by the steady state velocity distribution is identical to those obeyed by the distribution  
in the pseudo-steady state of the Maxwell gas with 
diffusive driving ($r_w=-1$)~\cite{Prasad:13}.

\section{Steady state Velocity Distribution}
\label{steadystate general gaussian}

 We use two diagnostic tools to obtain the tail of the steady state velocity distribution:
(1) by directly studying the characteristic function of the velocity distribution and
(2) by determining the ratios of large moments of the velocity distribution.

In the steady state, due to collisions being random, 
there are no correlations between velocities of two different particles in the thermodynamic limit. 
We note that for finite systems, there are  correlations that are proportional to $N^{-1}$~\cite{Prasad:13}.
The two point
joint probability distributions  can thus be written as a product of one-point probability distributions.
It is then straightforward to write
\begin{widetext}
\bea
P(v,t+1)=p \!\!\int \!\!\!\!\int \!\!dv_1  dv_2 P(v_1,t) P(v_2,t) \delta\left[\frac{1-r}{2} v_1 + \frac{1+r}{2} v_2 -v\right]+
(1-p) \!\!\int\!\! \!\!\int \!\!d\eta  dv_1 \Phi(\eta) P(v_1,t) \delta\left[\eta -r_w v_1 -v \right],
\label{eq:evolution}
\eea
\end{widetext}
where the first term on the right hand side describes the evolution due to collisions between particles
and the second term describes the evolution due to collision between particles and wall.
In the steady state, the velocity distributions become time independent and we use the
notation $\lim_{t\to \infty} P(v,t) = P(v)$. \Eref{eq:evolution} is best analyzed in the Fourier space.
Let  the characteristic function of the velocity distribution be defined as
\bea
Z(\lambda)=\langle\exp(-i\lambda v)\rangle.
\eea 
It can be shown from \eref{eq:evolution} that $Z(\lambda)$ satisfies the relation~\cite{Prasad:13}
\begin{equation}
Z(\lambda)=p Z\left(\frac{[1-r] \lambda}{2}\right)Z\left(\frac{[1+r] \lambda}{2}\right) + (1-p)
Z(r_w \lambda)f(\lambda),
\label{Z-continuous}
\end{equation}
where  $f(\lambda)\equiv\langle\exp(-i\lambda \eta)\rangle_\eta$. 
Equation~(\ref{Z-continuous}) is non-linear and non-local (in the argument of $Z$) and is not solvable in general. But it is possible to numerically
obtain the probability distribution for certain choices of the parameters.

When $r=0$ and $r_w=1/2$, \eref{Z-continuous}
takes the form,
 \begin{equation}
Z(\lambda)=p \left[Z\left(\frac{\lambda}{2}\right)\right]^2 + (1-p)Z\left(\frac{\lambda}{2}\right)f(\lambda), ~r=0, r_w=\frac{1}{2}.
\label{mgf steady state eqn special epsilon1/2 rw1/2}
\end{equation}
Thus, $Z(\lambda)$ is determined if
$Z(\lambda/2)$ is known. By iterating to smaller $\lambda$, and considering the initial value 
$Z(\lambda)=1-\lambda^2\langle v^2\rangle/2$
for small $\lambda$,  one can use this recursion relation to calculate characteristic function for any value of $\lambda$.
Here $\langle v^2\rangle$ may be calculated exactly [see \eref{steady state variance large N}].
The velocity distribution may be obtained from the inverse Fourier  transform of  $Z(\lambda)$. 

When $r_w=1$, \eref{Z-continuous} allows the tail statistics of $P(v)$ to be
determined exactly. In this case, the characteristic function satisfies the
relation
\begin{equation}
Z(\lambda)=\frac{ p
Z\left([1-r]\lambda/2\right)Z\left([1+r] \lambda/2 \right)}{[1-(1-p)
f(\lambda)]},~~ r_w=1.
\label{mgf steady state eqn for rw=1}
\end{equation}
Equation (\ref{mgf steady state eqn for rw=1}) may be iteratively solved to obtain
an infinite product involving simple poles. The behavior of the velocity 
distribution for asymptotically large velocities is determined by the pole
closest to the origin, and has the form $P(v)\sim \exp(-\lambda^*|v|)$,
where $\lambda^*$ is determined from 
$1-(1-p)f(\lambda)=0$~\cite{Prasad:13}.
When $r=1/2$, the iterative numerical scheme discussed above for dissipative
driving may be followed for determining the
characteristic function for the diffusive case.

The dynamics [Eqs.(\ref{interparticle collision},\ref{driving sec2})] also allows the calculation of the moments of the steady state distribution.
For the Maxwell model, it was shown that the equations for the two point correlation functions 
close~\cite{Prasad:13,Prasad:14}.
The closure can be also extended to one dimensional pseudo Maxwell models 
where  particles  collide only with nearest neighbor particles with equal rates~\cite{Prasad:16}.
Using this simplifying property, the variance of the steady state velocity distribution in the thermodynamic limit
was determined to be:
\begin{equation}
\langle v^2\rangle=
\frac{2\kappa\sigma^2}{1-r^2+2\kappa(1-r_w^2)},
\label{steady state variance large N}
\end{equation}
where $\kappa=(1-p)/p$ and $\sigma^2$ is the variance of the noise distribution.
On the other hand, the two-point velocity correlations in the steady state vanishes in the 
thermodynamic limit.

Among the higher moments,  the odd moments vanish as the velocity distributions is even. 
Define $2n$-th moment of the distribution to be $\langle v^{2n}\rangle=M_{2n}$. 
The evolution equation for $M_{2n}$ may be obtained by multiplying
\eref{eq:evolution} by $v^{2n}$, and integrating over the velocities. 
It  is then straightforward to show 
that they satisfy a recurrence relation,
\begin{align}
\label{moment ratio}
\Bigl[1-&\epsilon^{2n} - (1-\epsilon)^{2n} +\kappa \bigl(1-r_w^{2n}\bigr)\Bigr]\,
M_{2n}=
\notag \\
&\sum_{m=1}^{n-1}\binom{2n}{2m}\epsilon^{2m}(1-\epsilon)^{2n-2m}M_{2m}M_{2n-2m}
\notag
\\
&+\kappa\sum_{m=0}^{n-1}\binom{2n}{2m} r_w^{2m} M_{2m}
N_{2n-2m},
\end{align} 
where $\epsilon=(1-r)/2$ and $N_{i}$ is the $i$-th moment of the noise distribution.
\Eref{moment ratio} expresses $M_{2n}$ in terms of lower order moments.
Since $P(v)$ is a normalizable distribution, $M_0=1$. Also $M_2$ is given by \eref{steady state variance large N}. Knowing
these two moments, all higher order moments may be derived recursively using \eref{moment ratio}.

The ratios of moments may be used for determining the tail of the velocity distribution. 
Suppose the velocity distribution is a stretched exponential:
\bea
P(v)=\frac{b^{1/\beta}}{2\Gamma\left(1+\beta^{-1}\right)}\exp(-b|v|^\beta), ~
b,\beta>0,
\label{ansatz velocity pdf}
\eea
where $\Gamma$ is the Gamma function.
For this distribution the $2n^{th}$ moment is
\bea
 M_{2n}=b^{-2 n/\beta} \frac{\Gamma(\frac{2n+1}{\beta})}{\beta \Gamma(1+\frac{1}{\beta})},
\label{exact moments for stretched exp pdf}
\eea
such that that the ratios for large $n$ is
\bea
\label{ratio1}
\frac{ M_{2n}}{M_{2n-2}}\approx\left(\frac{2n}{b \beta}\right)^{2/\beta}, ~~n\gg1. 
\eea
Though \eref{ratio1} has been derived for the specific distribution given in \eref{ansatz velocity pdf}, the moment ratios will
asymptotically obey \eref{ratio1} even if only the tail of the distribution is a stretched exponential. This is because
large moments are determined only by the tail of the distribution. 
Thus, the exponent $\beta$ can be obtained unambiguously from
the asymptotic behavior of the moment ratios.

\subsection{Dissipative Driving ($r_w<1$) }
\label{dissipative driving}

\begin{figure}
\includegraphics[width=\hsize]{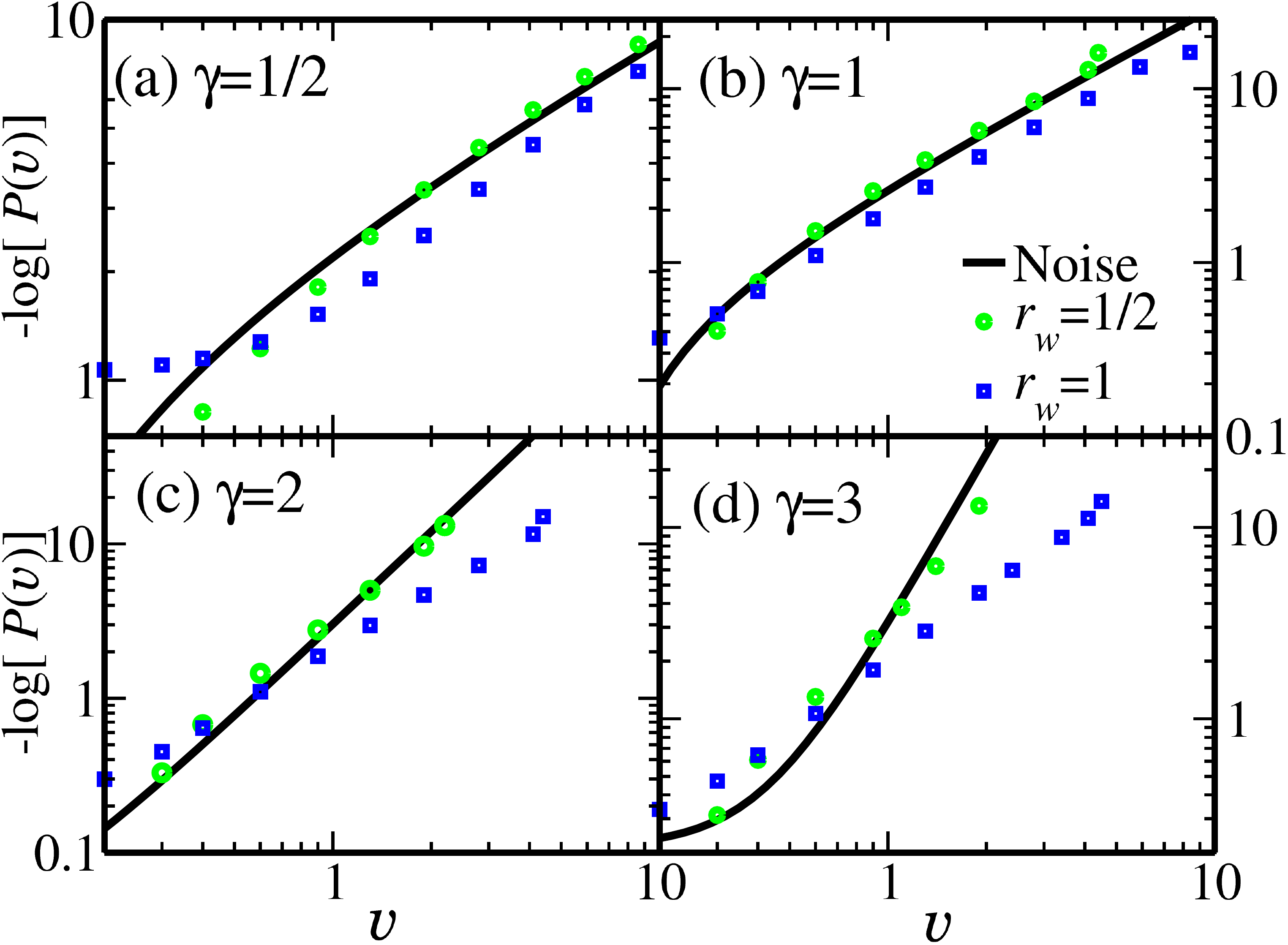}
\caption{(color online) The numerically calculated velocity distribution $P(v)$, obtained from the inverse
Fourier Transform of the characteristic
function $Z(\lambda)$,  for different noise distributions as described in  \eref{ansatz noise pdf} with (a) $\gamma=1/2$, (b) $\gamma=1$, (c) $\gamma=2$, and (d) $\gamma=3$ for $a=3$. $P(v)$ is computed for
$r_w=1/2$ (dissipative driving) and $r_w=1$ (diffusive driving) and compared with the noise distribution.}
\label{Fig1}
\end{figure}

We first evaluate the velocity distribution numerically by inverting the characteristic function $Z(\lambda)$.
For this calculation, $f(\lambda)$, the Fourier transform of the noise distribution in \eref{ansatz noise pdf}, is 
determined numerically using \eref{mgf steady state eqn special epsilon1/2 rw1/2}. \Fref{Fig1} shows the velocity distributions obtained for $\gamma=1/2,1,2,3$ for fixed $a=3$
(see \eref{ansatz noise pdf} for definition of $a$).
For the case  $r_w=1/2$, corresponding to dissipative driving,  the velocity distribution
$P(v)$  approaches the noise distribution for large velocities for all values of $\gamma$. This suggests
that the tail of the distribution is determined by the characteristics of the noise. However, using this method, it is not 
possible to extend the range of $v$ to larger values so that the large $v$ behavior may be determined unambiguously. 
The range of $v$ is limited by the 
precision to which   $f(\lambda)$ can be determined numerically. 

\begin{figure}
\includegraphics[width=\hsize]{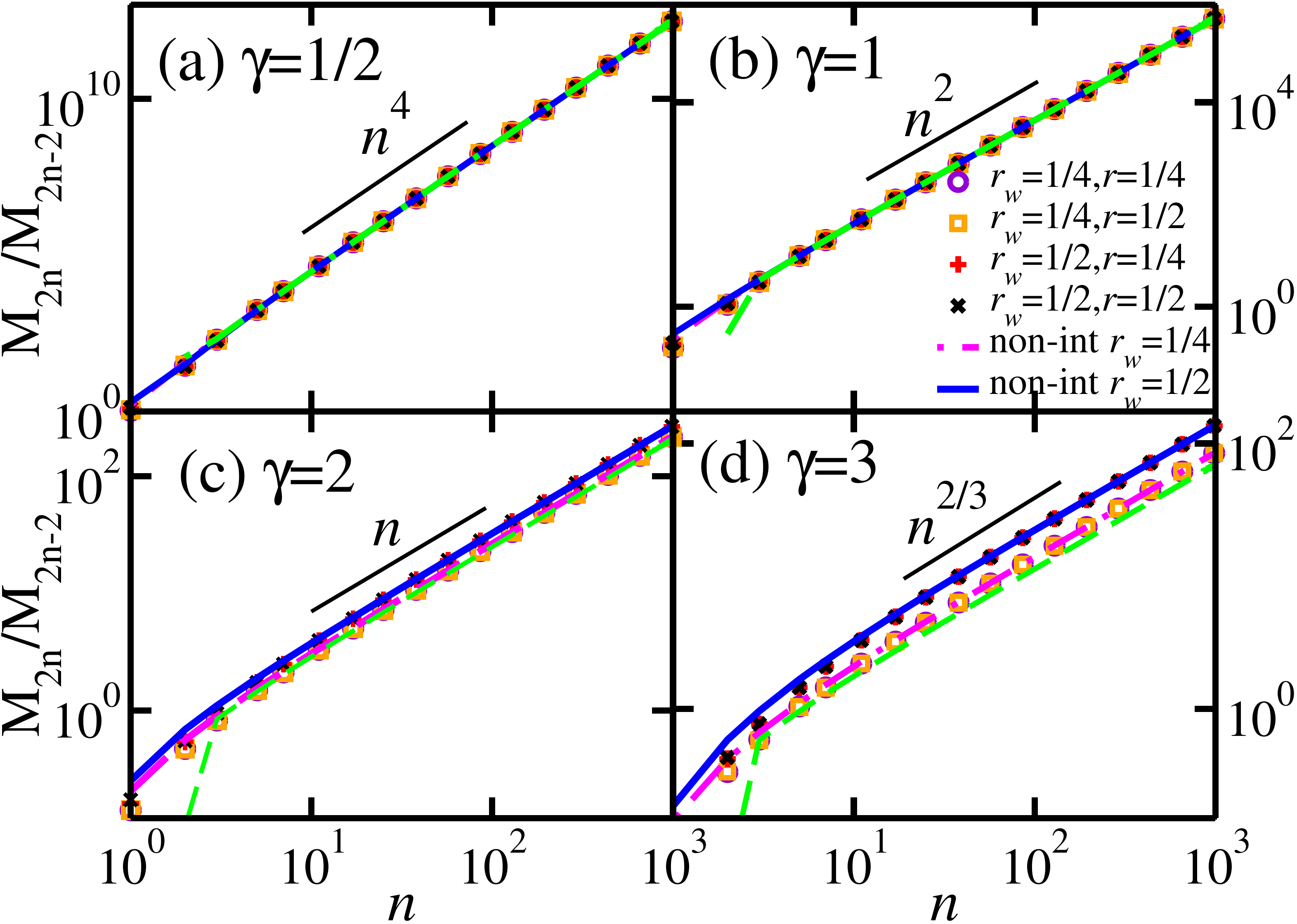}
\caption{(Color Online) The moment ratios  [see \eref{ratio1}] for  different noise distributions as described in  
\eref{ansatz noise pdf} with (a) $\gamma=1/2$, (b) $\gamma=1$, (c) $\gamma=2$, and (d) $\gamma=3$ for $a=3$. 
In each figure the ratios are plotted for $r=1/4,~1/2$, as well as $r_w=1/4,~1/2$, corresponding to dissipative driving.
These are is compared with the moment ratios of the  non-interacting system in which collisions are  ignored, as well as the 
noise distribution (dashed green line).}
\label{Fig2}
\end{figure}

The ratios of moments [see \eref{ratio1}] is a more robust method for determining the tail of the velocity distribution.
The moments are calculated from  the recurrence relation \eref{moment ratio} where the moments of the noise distribution
described in \eref{ansatz noise pdf} is given by
\bea
N_{2n}=a^{-2 n/\gamma} \frac{\Gamma(\frac{2n+1}{\gamma})}{\gamma \Gamma(1+\frac{1}{\gamma})}.
\eea
The numerically obtained moment ratios of the steady state velocity distribution
for dissipative driving is shown in \fref{Fig2}, for  different noise distributions characterized by $\gamma$.
The moment ratios increase with $n$ as a power law with an exponent $2/\gamma$, independent of the
value of $r_w$ and the coefficient of restitution $r$. Comparing with \eref{ratio1}, we obtain $\beta=\gamma$,
and that the tail of the velocity distribution is determined by the noise distribution. We also compare the results with
those for driven non-interacting particles. Here, collisions between particles are completely ignored so that the time
evolution of particles are independent of each other, and each particle is driven
independently. For the range of parameters, considered, the moment ratios of the interacting system is asymptotically
indistinguishable from that of the non-interacting system, showing that for dissipative driving collisions between particles
do not affect the tails of the velocity distribution. The moment ratios are also compared with those of the noise
distribution. Here, we observe that while the ratios have the same power law exponent, the prefactor is different.

\begin{figure}
\includegraphics[width=1\hsize]{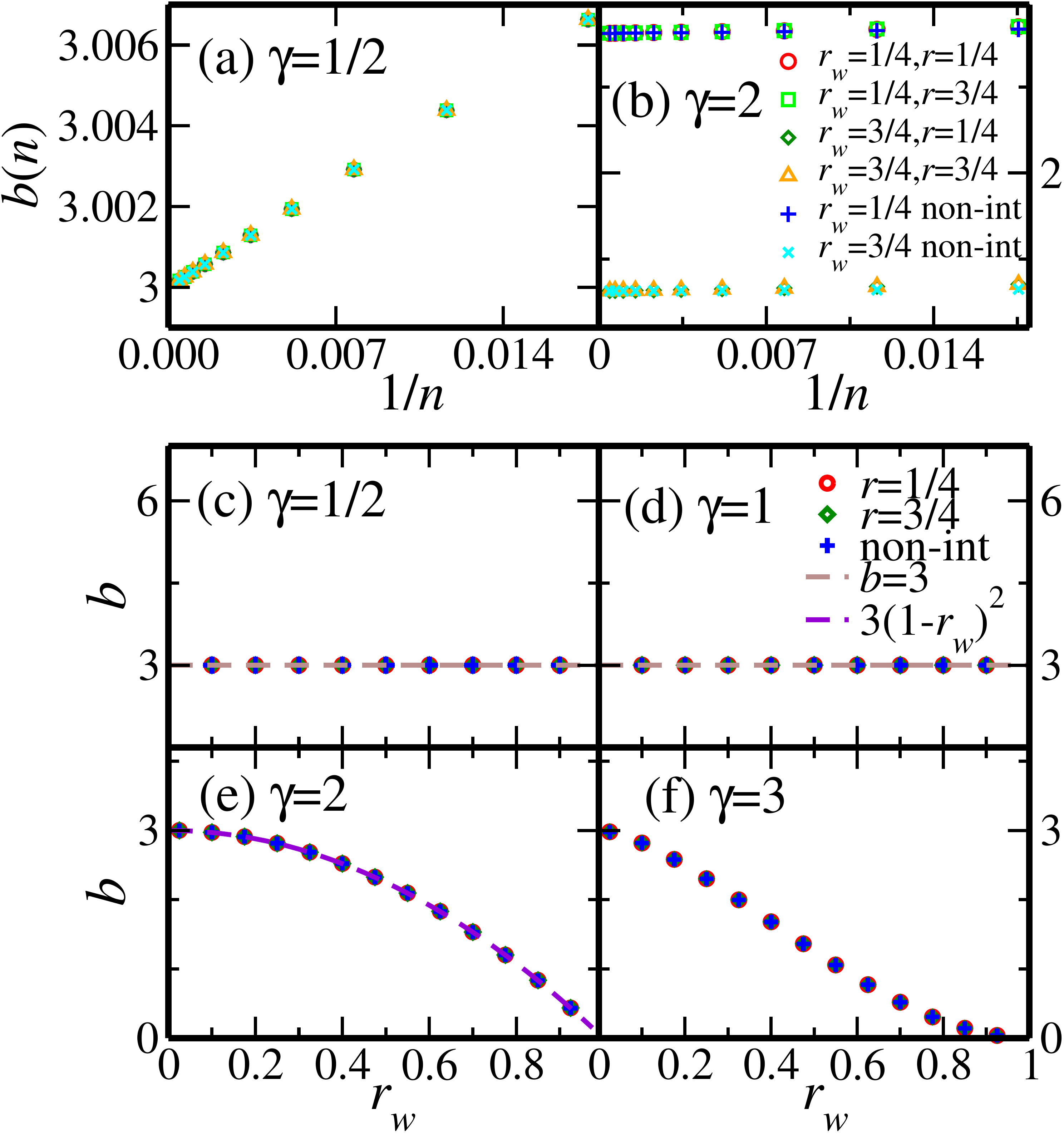}
\caption{(Color Online) The coefficient $b(n)$ obtained from \eref{b from ratio}
for (a) $\gamma=1/2$, (b) $\gamma=2$ varies linearly with $n^{-1}$ for dissipative driving $r_w<1$. The choice of $r$
and $r_w$ are the same in both plots and labeled in (b). The corresponding $b(n)$ obtained for the non-interacting
system are also shown. The variation of $b=b(\infty)$ with $r_w$ is shown for (c) $\gamma=1/2$, (d) $\gamma=1$,
(e) $\gamma=2$ and (f) $\gamma=3$.
}
\label{dissipative driving coeff b for all gamma 1}
\end{figure}

We now determine the constant $b$ in the exponential in \eref{ansatz velocity pdf}. It may be 
determined from \eref{ratio1} once $\beta$ is determined.
Rearranging \eref{ratio1}, we obtain
\bea
\label{b from ratio}
b(n)\approx\frac{2n}{\beta}\left(\frac{ M_{2n}}{M_{2n-2}}\right)^{-\beta/2}, ~~n\gg1. 
\eea
Figures~\ref{dissipative driving coeff b for all gamma 1}~(a) and(b) show the variation of $b(n)$ with $n$ for different $\gamma$. 
We find that for large $n$,
$b(n)$ is independent of coefficient of restitution $r$, but may depend on $r_w$. Also, we find that $b-b(n)\propto n^{-1}$ for 
all values of $\gamma$.
Figures~\ref{dissipative driving coeff b for all gamma 1}~(c) and(f) show the variation of $b$ with $r_w$ for different $\gamma$.
For $\gamma=1/2$ and $1$, $b$ is independent of $r_w$, while for $\gamma=2$ and $3$, it depends on $r_w$. We have checked 
that $b$ is independent of $r_w$ for $\gamma$ up to $1$. In Figures~\ref{dissipative driving coeff b for all gamma 1}~(c) and(f), the
values of $b$ are also compared with that obtained for a non-interacting system in which collisions between particles
are ignored. We find that the values of $b$ for both the interacting and non-interacting system coincide. In addition, for $\gamma\leq 1$,
we find that  the value of $b$ approaches the value $a$ characterizing  the noise distribution. 

\subsection{Velocity distributions for diffusive driving}
\label{diffusive driving}

The Maxwell gas with diffusive driving ($r_w=-1$) does not have a 
steady state in the long time limit, when the total energy diverges.  However, it has
a pseudo steady state solution that is valid at intermediate times. 
On the other hand when $r_w=1$ the system reaches a steady state at large time.
It has been shown that the velocity distribution in the pseudo steady state for the case $r_w=-1$ is the same as the velocity
distribution in the steady state of the system with 
$r_w=1$~\cite{Prasad:13}.
For $r_w=1 $ and  $\eta$ taken from a Gaussian distribution, the
velocity distribution was shown to have an exponential distribution~\cite{Prasad:13}.
In this section, we determine this steady state for other noise distributions.

In \fref{Fig1}, the numerically obtained $P(v)$ is shown for different values of $\gamma$.
We find that for $\gamma=1/2,1$ the velocity distribution approaches the noise distribution. 
Interestingly, when $\gamma=2,3$ the velocity distribution deviates significantly from the noise distribution. 
While the data for $\ln P(v)$  appears to vary linearly with $v$, the range is limited and it is not possible
to unambiguously conclude that $P(v)$ is exponential independent of the noise distribution.

\begin{figure}
\includegraphics[width=\hsize]{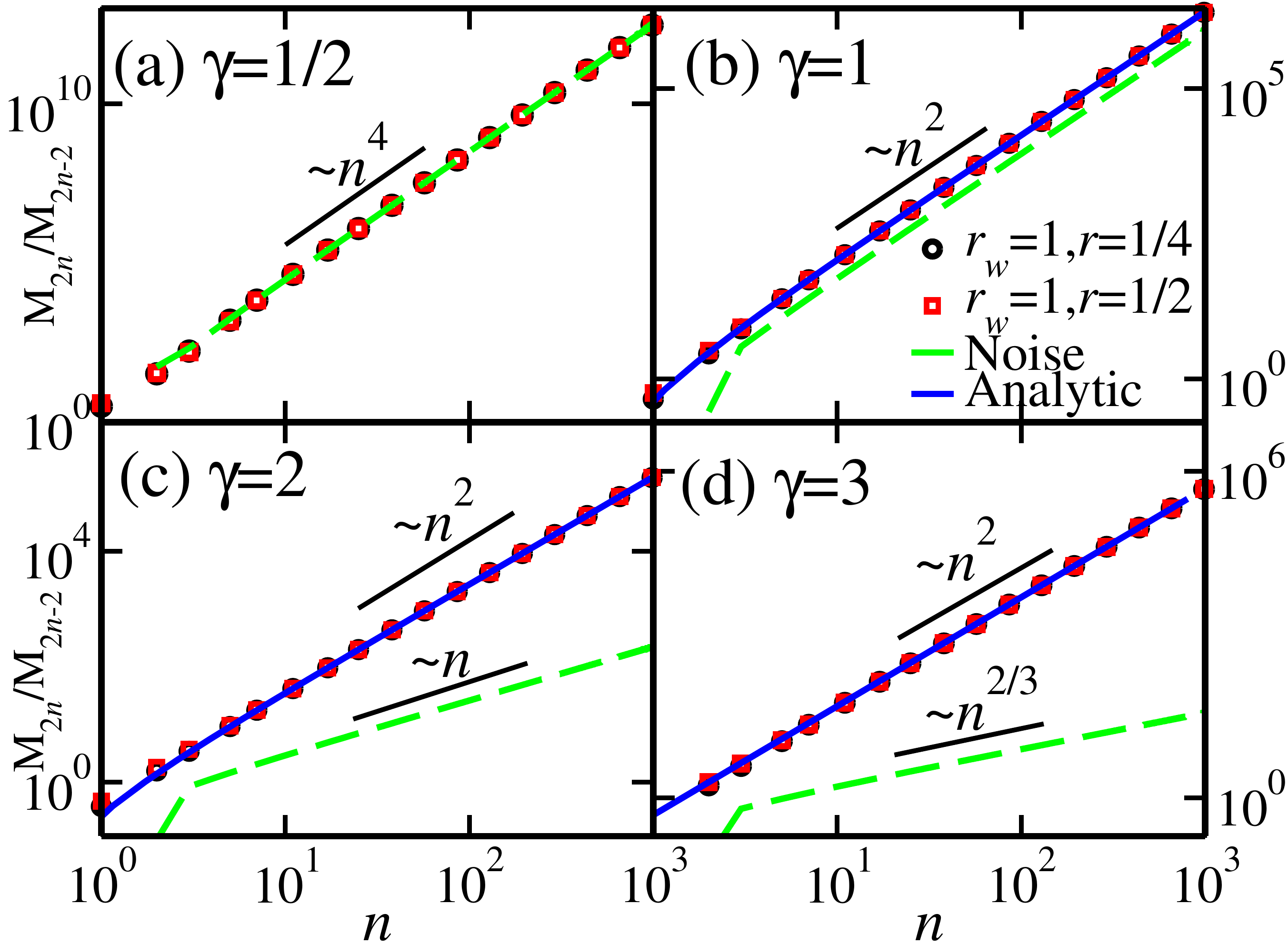}
\caption{(color online) The moment ratios  [see \eref{ratio1}] for  different noise distributions as described in  
\eref{ansatz noise pdf} with (a) $\gamma=1/2$, (b) $\gamma=1$, (c) $\gamma=2$, and (d) $\gamma=3$ for $a=3$. The
data are for $r_w=1$ (diffusive driving) and the ratios are plotted for $r=1/4$,and $1/2$.
These are  compared with the noise distribution (dashed green line). In (b), (c) and (d), we also plot moment ratios 
for the exponential distribution with analytically obtained  value of $\lambda^*$ [See Eqs. (\ref{eq:lambdastar1}), (\ref{eq:lambdastar2})].}
\label{Fig3}
\end{figure}

As for the dissipative case, the better tool to probe the tail of the distributions is the moment ratios \eref{ratio1}.
\Fref{Fig3} shows that moment ratios increase with $n$ as a power law. The power law exponent is
$2/\gamma$ for $\gamma<1$ [see \fref{Fig3}(a)] and equal to 2 for $\gamma \geq 1$ [see \fref{Fig3}((b)-(d)].
Thus, we conclude that $\beta=\min[\gamma,1]$. Thus, $P(v)$ is universal, and has an exponential tail
for $\gamma \geq 1$.

The exact form of the universal exponential tail can be analytically obtained as follows.
If the velocity distribution has the form  $P(v)=(\lambda^*/2)\exp(-\lambda^*|v|)$,
the moment ratios in the large $n$ limit behaves as $M_{2n}/M_{2n-2}\approx (4n^2-2n)/(\lambda^*)^2$. 
But we have seen in \sref{steadystate general gaussian} that, for diffusive driving
\eref{mgf steady state eqn for rw=1} satisfies  a solution such that the velocity distribution
is determined by the pole nearest to the origin $\pm i\lambda^*$  obtained from relation $1=(1-p)f(\lambda)$.
When $\gamma=1$, the pole has the form given by 
\bea
\label{eq:lambdastar1}
\lambda^*&=&\pm a \sqrt{p}, ~\gamma=1,\\
\label{eq:lambdastar2}
\lambda^*&=&\pm\frac{\sqrt{-2\ln(1-p)}}{ \sigma}, ~\gamma=2.
\eea 
When $\gamma=3$, we obtain complicated Hypergeometric function for $f(\lambda)$ from which 
$\lambda^*$ may be determined numerically.
The moment ratios thus obtained  are
plotted in \fref{Fig3}(b), (c), and (d)  which matches with the numerically calculated moment ratio. 
It can be seen that when $\gamma<1$, there is no  $\lambda^*$
which satisfies the relation $1=(1-p)f(\lambda)$.
 
\begin{figure}
\includegraphics[width=1\hsize]{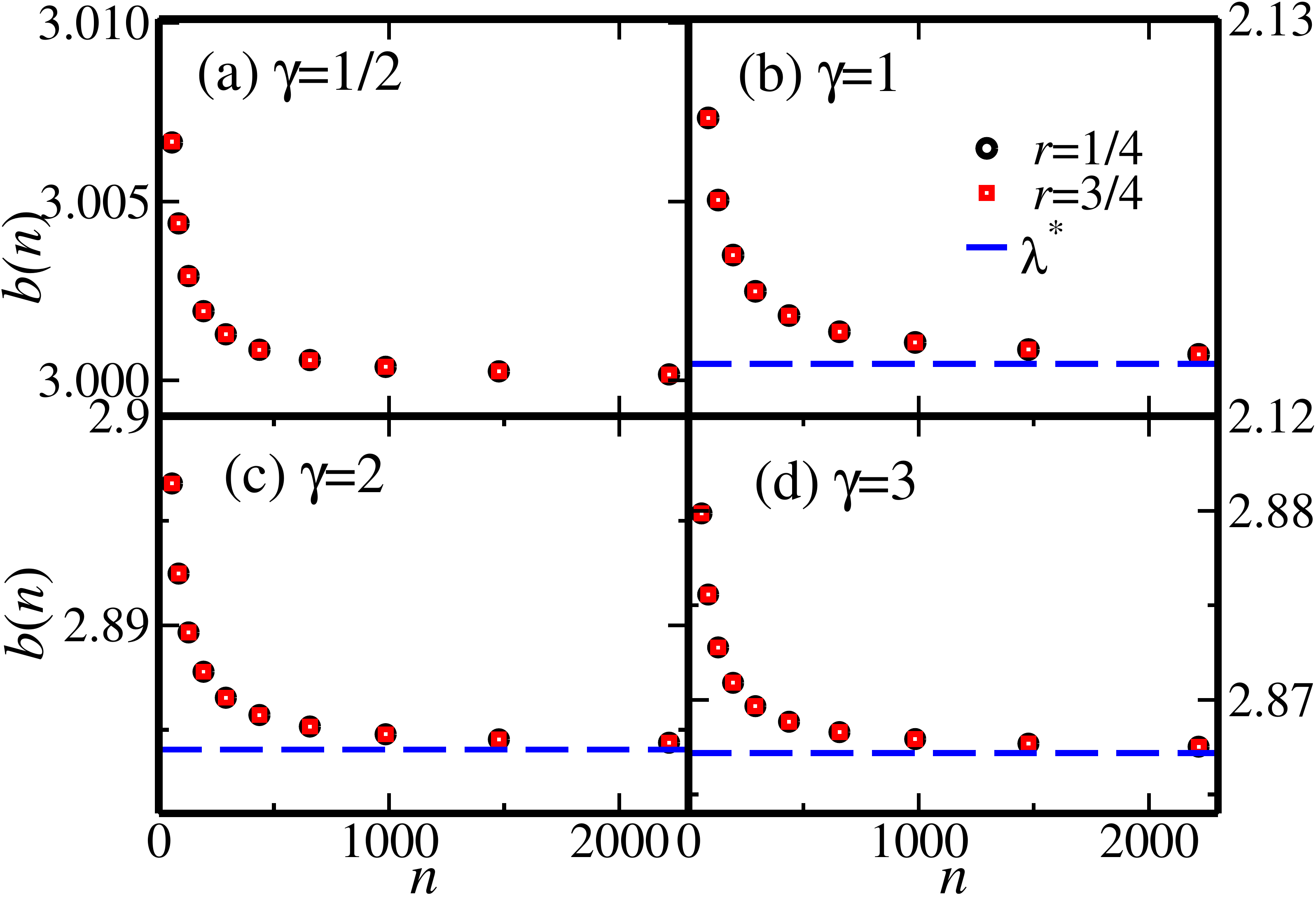}
\caption{(Color Online) The variation of the coefficient $b(n)$ with $n$ obtained from \eref{b from ratio}
 for diffusive driving $r_w=1$ and for different values of $r$,
for different noise distribution characterized by (a)~$\gamma=1/2$, (b)~$\gamma=1$,
 (c)~$\gamma=2$, (d)~$\gamma=3$. The dashed line corresponds to the analytically obtained asymptotic value 
 $\lambda^*$ [See Eqs. (\ref{eq:lambdastar1}), (\ref{eq:lambdastar2})].}
\label{diffusive driving coeff b for gamma}
\end{figure}

From \eref{b from ratio}, we obtain the coefficient $b$ 
for the diffusively driven system and is shown in \fref{diffusive driving coeff b for gamma}.
It is seen that when $\gamma<1$,  the coefficient $b(n)$ approaches that of the noise distribution $a=3$.
For  $\gamma\ge1$, $b$ is calculated by substituting $\beta=1$ in \eref{b from ratio}. One finds  
in this case that $b$ approaches $\lambda^*$ which is  obtained analytically.

\section{Non-interacting system}
\label{section non-interacting}

We showed in \sref{dissipative driving} that, for dissipative driving, the tail of the velocity distribution $P(v)$ is identical to that
of a non-interacting system in which collisions between particles may be ignored. In this section, we determine
the velocity distribution of the non-interacting system in terms of the noise distribution. 
In the non-interacting system, the particle is driven at each time step. If $v_n$ is the velocity after the $n^{th}$ collision, then
\bea
v_n=-r_w v_{n-1}+\eta_{n-1}.
\eea
For a particle that is initially at rest ($v_0=0$), 
\bea
v_n=\displaystyle\sum_{m=0}^{n-1}r_w^m \eta_{n-m-1}=\displaystyle\sum_{m=0}^{n-1}r_w^m \eta_{m},
\eea
where the second equality is in the statistical sense, and follows from the fact that noise is uncorrelated and
therefore the order is irrelevant.

Now, consider the moment generating function of the noise distribution, 
$\langle\exp(-\lambda\eta)\rangle\equiv\exp[\mu(\lambda)]$
where $\mu(\lambda)$ is the cumulant generating function,
\bea
\mu(\lambda)\equiv\displaystyle\sum_{i=1}^{\infty}\frac{\lambda^{2n}}{2n!}C_{2n},
\label{cgf noise dfn}
\eea
where $C_{2n}$ is the $2n^{\text{th}}$ cumulant of the noise distribution.
It has been assumed that the noise distribution is symmetric such that only even cumulants are non-zero.
The moment generating function of the velocity after infinite time-steps is,
\begin{align}
\langle\exp(-\lambda v_\infty)\rangle_\eta &=& \left\langle\exp\left[-\lambda \sum_{m=0}^{\infty}r_w^{m} \eta_{m}\right]\right\rangle_\eta ,\nn\\
&=&\exp\left[-\sum_{m=0}^{\infty}\mu(r_w^{m}\lambda )\right].
\end{align}
From the definition of $\mu(\lambda)$ [see \eref{cgf noise dfn}], we obtain
\bea
\mu(r_w^{m}\lambda)=\displaystyle\sum_{n=1}^{\infty}\frac{(r_w^{m}\lambda)^{2n}}{2n!}C_{2n}.
\label{cgf vel dfn}
\eea

Summing over $m$,
\begin{align}
\sum_{m=0}^{\infty}\mu(r_w^{2m}\lambda)&=&\sum_{m=0}^{\infty}\sum_{n=1}^{\infty}\frac{(r_w^{m}\lambda)^{2n}}{2n!}C_{2n},\nn\\
&=&\sum_{n=1}^{\infty}\frac{\lambda^{2n}}{2n!}\left(\frac{1}{1-r_w^{2n}}\right) C_{2n}.
\label{cgf}
\end{align}
But, $\langle\exp(-\lambda v_\infty)\rangle= \exp\left[\xi(\lambda)\right]$ where $\xi(\lambda)$ 
is the cumulant generating function of the velocity distribution at large times,
\begin{align}
\xi(\lambda)=\displaystyle\sum_{n=1}^{\infty}\frac{\lambda^{2n}}{2n!} D_{2n},
\end{align}
where $D_{2n}$ is the $2n^{\text{th}}$ cumulant of the velocity distribution.
Comparing  with \eref{cgf}, we obtain
\begin{align}
D_{2n}=\frac{C_{2n}}{1-r_w^{2n}}.
\end{align} 
For large $n$, behavior  of the cumulants of the velocity distribution 
approaches that of the noise distribution. Thus, by knowing all cumulants,
the velocity distribution of the non-interacting system is completely 
determined.

\section{Discussion and conclusion}
\label{conclusion}

\begin{figure}
\includegraphics[width=\hsize]{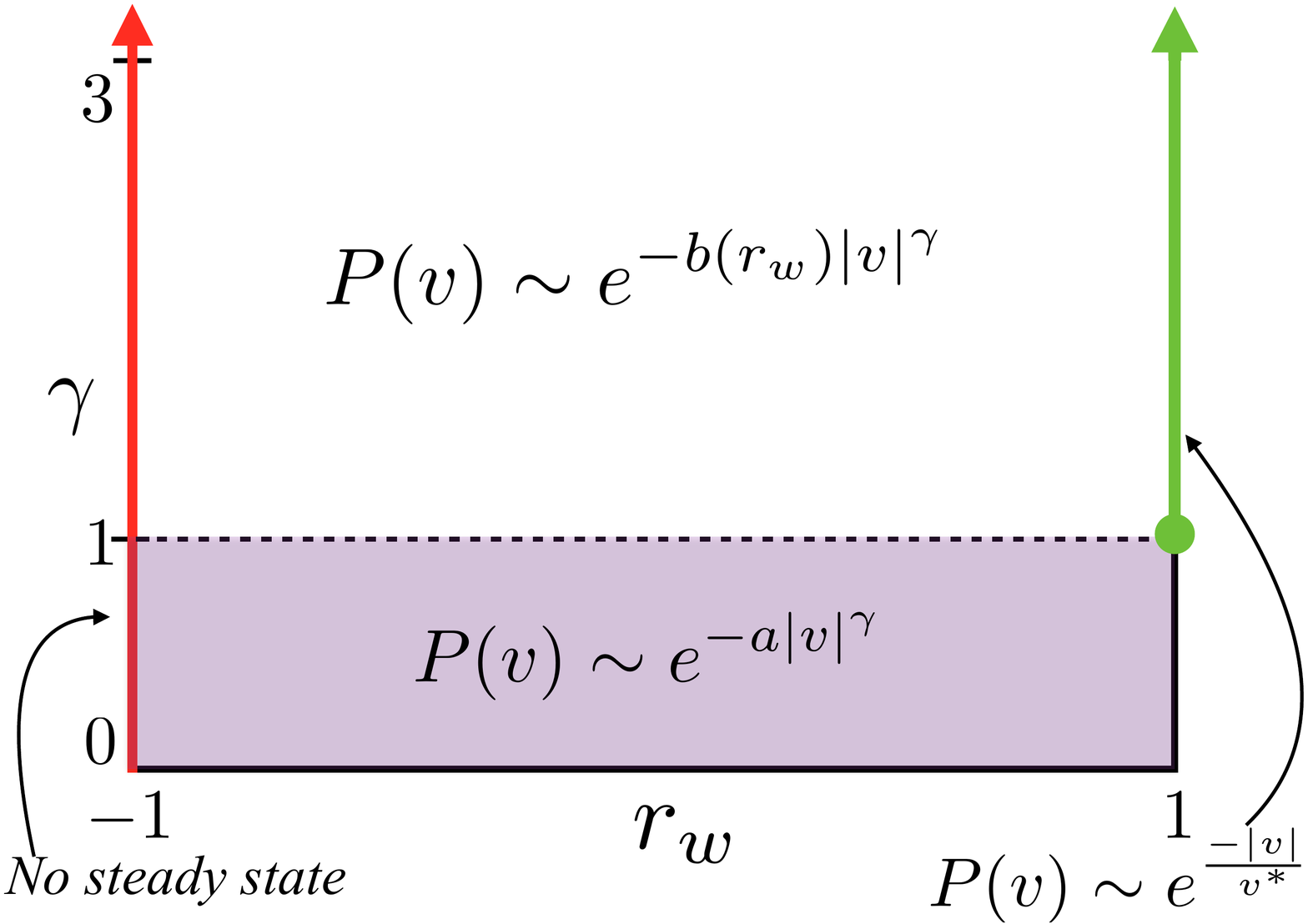}
\caption{(color online) Schematic diagram summarizing the results obtained in paper. The parameters 
$r_w\in[-1,1]$ is the coefficient of restitution of wall-particle collisions and $\gamma$ characterizes the noise distribution 
[see \eref{ansatz noise pdf}]. When $r_w=-1$, the system does not reach a time independent steady state.
When $r_w=1$, $P(v)$ is universal when $\gamma \geq 1$, and has the same asymptotic behavior as
the noise distribution when $\gamma<1$.
When the driving is dissipative ($|r_w|<1$), $P(v)$ has the same asymptotic behavior as the noise
distribution for $\gamma <1$. When $\gamma \geq 1$, the coefficient in the exponential gets modified.
}
\label{Fig6}
\end{figure}

In summary, we considered an inelastic one component Maxwell gas in which 
particles are driven through collisions with a wall. We determined precisely the tail
of the velocity distribution $P(v)$ by analyzing the asymptotic behavior of the ratio of consecutive moments. 
Our main results are: (1) For dissipative driving, the tail of $P(v)$ is identical to that of 
the corresponding non-interacting system where collisions are ignored. By solving the non-interacting 
problem, the cumulants of the velocity distribution may be expressed in terms of the noise distribution
and. Thus, $P(v)$  is highly non-universal. (2) For diffusive driving, $P(v)$ is universal and decays exponentially 
when the noise distribution decays faster than exponential. If $\Phi(\eta)$ decays slower than exponential,
then $P(v)$ is non-universal and the tails are similar to the tail of $\Phi(\eta)$. These results are summarized in
\fref{Fig6}.

These results generalize the results in Ref.~\cite{Prasad:13}, where it was shown that for dissipative driving that when
the noise distribution
is gaussian or Cauchy, the tails of the velocity distribution are similar to that of the noise distribution. 
The results are also consistent with the intuitive understanding that the tails
of velocity distribution are bounded from below by the noise distribution. 
This is because the tails are populated by particles that have been recently driven and then do not undergo any
collision.
We expect that more complicated kernels of collision will not change the result. 
This could be the reason why many of the experimental results~\cite{Blair:01} see non-universal behavior.
However, there are experiments that see universal behavior~\cite{Losert:99,Rouyer:00}. In these experiments the $P(v)$ 
is measured in directions perpendicular to the driving direction. It may be that the details of the driving are 
lost when energy is transferred to other directions. Transferring energy in other directions ensures that collisions
cannot be ignored, unlike the case of one-component Maxwell gas studied in this paper. 
The two component Maxwell model is a good starting point to 
answer this question. Methods developed in the paper will be useful to analyze the same. This is a 
promising area for future study.  

\acknowledgments
This research was supported in part by the International Centre for Theoretical Sciences (ICTS) during a visit for participating in the program -Indian Statistical Physics Community Meeting 2016 (Code: ICTS/Prog-ISPC/2016/02)


\end{document}